\documentclass[12pt]{article}

\usepackage{scicite}

\usepackage{times}
\usepackage{graphicx}
\usepackage{xspace}
\usepackage{siunitx}
\usepackage{authblk}
\usepackage{xr}
\usepackage{rotating, longtable}

\newcommand{\apj}{Astrophys. J. }

\newcommand{\aap}{Astron. Astrophys.}
\newcommand{\apjl}{Astrophys. J. }
\newcommand{\mnras}{Mon. Not. R. Astron. Soc.}

\newcommand{\apjs}{Astrophys. J.}
\newcommand{\nat}{{\it Nature}}

\newcommand{\pasp}{Publ. Astron. Soc. Pac.}

\newcommand{\annrev}{Annu. Rev. Nuc. Part. Sci.}
\newcommand{\procspie}{Proceedings of the Society of Photo-Optical Instrumentation Engineers Conference Series}

\newcommand{\sn}{SN\xspace}
\newcommand{\sne}{SNe\xspace}
\newcommand{\snia}{SN~Ia\xspace}
\newcommand{\sneia}{SNe~Ia\xspace}

\newcommand{\be}{\begin{equation}}
\newcommand{\ee}{\end{equation}}
\newcommand{\om}{\Omega_M}

\newenvironment{sciabstract}{%
\begin{quote} \bf}
{\end{quote}}

\renewcommand\refname{References and Notes}

\newcounter{lastnote}

\title{iPTF16geu: A multiply imaged, gravitationally lensed type Ia supernova} 
\author { {A.~Goobar,$^{1\ast}$ R.~Amanullah,$^{1}$
    S.~R.~Kulkarni,$^{2}$ P.~E.~Nugent,$^{3,4}$J.~Johansson,$^{5}$}
  {C.~Steidel,$^{2}$ D.~Law,$^{6}$ E.~M\"ortsell,$^{1}$ R.~Quimby,$^{7,8}$
    N.~Blagorodnova,$^{2}$ A.~Brandeker,$^{9}$} {Y.~Cao,$^{10}$ A.~Cooray,$^{11}$ R.~Ferretti,$^{1}$ C.~Fremling,$^{12}$
    L.~Hangard$^{1}$, M.~Kasliwal,$^{2}$ T.~Kupfer,$^{2}$ R.~Lunnan,$^{2,9}$} {F.~Masci,$^{13}$
    A.~A.Miller,$^{15,16}$ H.~Nayyeri,$^{11}$ J.~D.~Neill,$^{2}$ E.~O.~Ofek,$^{5}$ S.~Papadogiannakis$^{1}$, T.~Petrushevska$^{1}$,
    V.~Ravi,$^{2}$} {J.~Sollerman,$^{12}$ M.~Sullivan,$^{14}$ F.~Taddia,$^{12}$ R.~Walters,$^{2}$ D.~Wilson,$^{11}$ L.~Yan,$^{2}$ O.~Yaron$^{5}$}
  \\
  \normalsize{$^{1}$The Oskar Klein Centre, Department of Physics,
    Stockholm University,
    Albanova University Center, SE 106 91 Stockholm, Sweden, } \\
  \normalsize{$^{2}$Cahill Center for Astrophysics, 
    California Institute of Technology, Pasadena, CA 91125, USA, } \\
  \normalsize{$^{3}$Department of Astronomy, 
    University of California, Berkeley, CA 94720-3411, USA, } \\
  \normalsize{$^{4}$Lawrence Berkeley National Laboratory, 
    1 Cyclotron Road, MS 50B-4206, Berkeley, CA 94720, USA, } \\
  \normalsize{$^{5}$Department of Particle Physics and Astrophysics, 
    Weizmann Institute of Science, Rehovot 7610001, Israel, } \\
  \normalsize{$^{6}$Space Telescope Science Institute, 
    3700 San Martin Drive, Baltimore, MD 21218, USA, }\\
  \normalsize{$^{7}$Department of Astronomy, San Diego State University, 
    San Diego, CA 92182, USA, } \\
  \normalsize{$^{8}$Kavli IPMU (WPI), UTIAS, The University of Tokyo,
    Kashiwa, Chiba 277-8583, Japan, } \\
  \normalsize{$^{9}$ Department of Astronomy, 
    Stockholm University, Albanova, SE 10691 Stockholm, Sweden, } \\
  \normalsize{$^{10}$eScience Institute and Department of Astronomy, University of Washington, 3910 15th Ave NE
    Seattle, WA 98195-1570, USA, }\\
  \normalsize{$^{11}$Department of Physics \& Astronomy, University of California, Irvine,  CA
92697}\\
  \normalsize{$^{12}$The Oskar Klein Centre, Department of Astronomy,
   Stockholm University,
   Albanova University Center, SE 106 91 Stockholm, Sweden,}\\
  \normalsize{$^{13}$Infrared Processing and Analysis Center, 
    California Institute of Technology, Pasadena, CA, 91125, USA,}\\
  \normalsize{$^{14}$Department of Physics and Astronomy, 
    University of Southampton, Southampton, SO17 1BJ, UK,} \\ 
   \normalsize{$^{15}$ Center for Interdisciplinary Exploration and Research in Astrophysics (CIERA) and Department of Physics and Astronomy, Northwestern University, 2145 Sheridan Road, Evanston, IL 60208, USA,}\\
   \normalsize{$^{16}$ The Adler Planetarium, 1300 S. Lakeshore Drive, Chicago, IL 60605, USA} \\

  \normalsize{$^\ast$To whom correspondence should be addressed;
    E-mail: ariel@fysik.su.se.}  }

\date{}

\def\lsim{\raise0.3ex\hbox{$<$}\kern-0.75em{\lower0.65ex\hbox{$\sim$}}}
\def\gsim{\raise0.3ex\hbox{$>$}\kern-0.75em{\lower0.65ex\hbox{$\sim$}}}
\def\arcsec{\hbox{$^{\hbox{\rlap{\hbox{\lower4pt\hbox{$\,\prime\prime$}}
          }\hbox{$\frown$}}}$}}

\def\arcmin{\hbox{$^{\hbox{\rlap{\hbox{\lower4pt\hbox{$\;\prime$}}
          }\hbox{$\frown$}}}$}}

\begin{document} 


\baselineskip24pt


\maketitle


\begin{sciabstract}
We report the discovery of a multiply-imaged gravitationally lensed Type Ia  supernova,  
iPTF16geu (SN 2016geu), at redshift $z=0.409$.  This phenomenon could be identified 
because the light from the stellar explosion was magnified more than fifty times by the curvature 
of space around matter in an intervening galaxy.
  
We used high spatial resolution observations to resolve four images of the
lensed supernova,  approximately \ang{;;0.3} from the center of the foreground
galaxy. The observations probe a physical scale of $\sim$1 kiloparsec, smaller  than what is typical in  other studies of extragalactic gravitational lensing. The large magnification and symmetric image configuration implies close alignment between the line-of-sight to the supernova and
the lens.  The relative magnifications of the four images provide evidence for sub-structures in the lensing galaxy. 
\end{sciabstract}

One of the foundations of Einstein's theory of General Relativity is that matter
  curves the surrounding space-time. For the rare cases of nearly perfect alignment between
  an astronomical source, an intervening massive object and the observer, 
  multiple images of a single source can be detected, 
 a phenomenon known as strong gravitational lensing.
  
 Although many strongly lensed galaxies and quasars have been detected
 to date, it has proved extremely difficult  to find multiply-imaged lensed  supernova (\sn) explosions. 
  Type Ia supernovae (\sneia) 
  are particularly interesting sources due to their
   ``standard candle'' nature.  These explosions have nearly identical peak luminosity
   which makes them excellent distance indicators in
 cosmology \cite{2011ARNPS..61..251G}. 
 For lensed \sneia, the standard candle property allows the flux magnification to be estimated directly, independent
 of any model related to the lensing galaxy \cite{Kolatt:1997zh,Oguri:2002ku}. This removes important 
 degeneracies in gravitational lensing measurements, the  mass-sheet degeneracy \cite{1985ApJ...289L...1F} 
 and the source-plane degeneracy \cite{2013A&A...559A..37S}.

   A lensed \snia  at redshift $z=1.388$ with a large amplification ($\mu\sim 30$) , PS1-10afx,  where multiple images could have been expected, has been reported earlier \cite{2013ApJ...768L..20Q}. A foreground lens was later identified at $z=1.117$ \cite{2014Sci...344..396Q}. However, at the time of the discovery several interpretations were discussed,
  including a super-luminous supernova \cite{2013ApJ...767..162C}. 
Since the lensed \snia hypothesis was only
  accepted long after the explosion had faded, no high spatial
  resolution imaging could be carried out in that case to verify the strong lensing nature of the system.
  Multiple-images  of  another supernova, \sn Refsdal \cite{2015Sci...347.1123K}, were discovered in 
   a  Hubble Space Telescope (HST) survey of the massive galaxy cluster MACS J1149.6+2223. As the source was identified as a core-collapse supernova it could not be used to measure the
  lensing magnification directly.


Thanks to the well-known characteristics of their time-dependent brightness in optical and near-infrared filters  (the \sn  lightcurves), multiply-imaged \sneia
  are also ideally suited to measure time-delays in the arrival of the
  images. This  provides a direct probe of the Hubble constant, the cosmological parameter measuring  the expansion rate of the universe\cite{1964MNRAS.128..307R}, as well as
  leverage for studies of dark energy 
  \cite{2002A&A...393...25G,2013ApJ...766...70S}, the cosmic constituent responsible for the accelerated expansion of the universe. 

The intermediate Palomar Transient Factory (iPTF) searches the sky for new transient phenomena at optical
wavelengths. It uses image differencing between repeated observations
\cite{cnk16} with a large field-of-view camera
(7.3 sq.deg) at the 48-inch telescope (P48) at the Palomar
Observatory \cite{2009PASP..121.1395L}.  
The first detection of iPTF16geu, with a statistical significance of five standard deviations (5$\sigma$), is from 2016 September 5. The new
source was first recognized by a human scanner on September 11  \cite {ATEL9603}.  iPTF16geu (also known as SN 2016geu) was found near the center of the galaxy
SDSS\,J$210415.89$-$062024.7$, at
right ascension $21^h$$4^m$$15.86^s$  and declination \ang{-06;20;24.5} (J2000).

\begin{figure}[!htb]
	\centering
	\includegraphics[width=\textwidth]{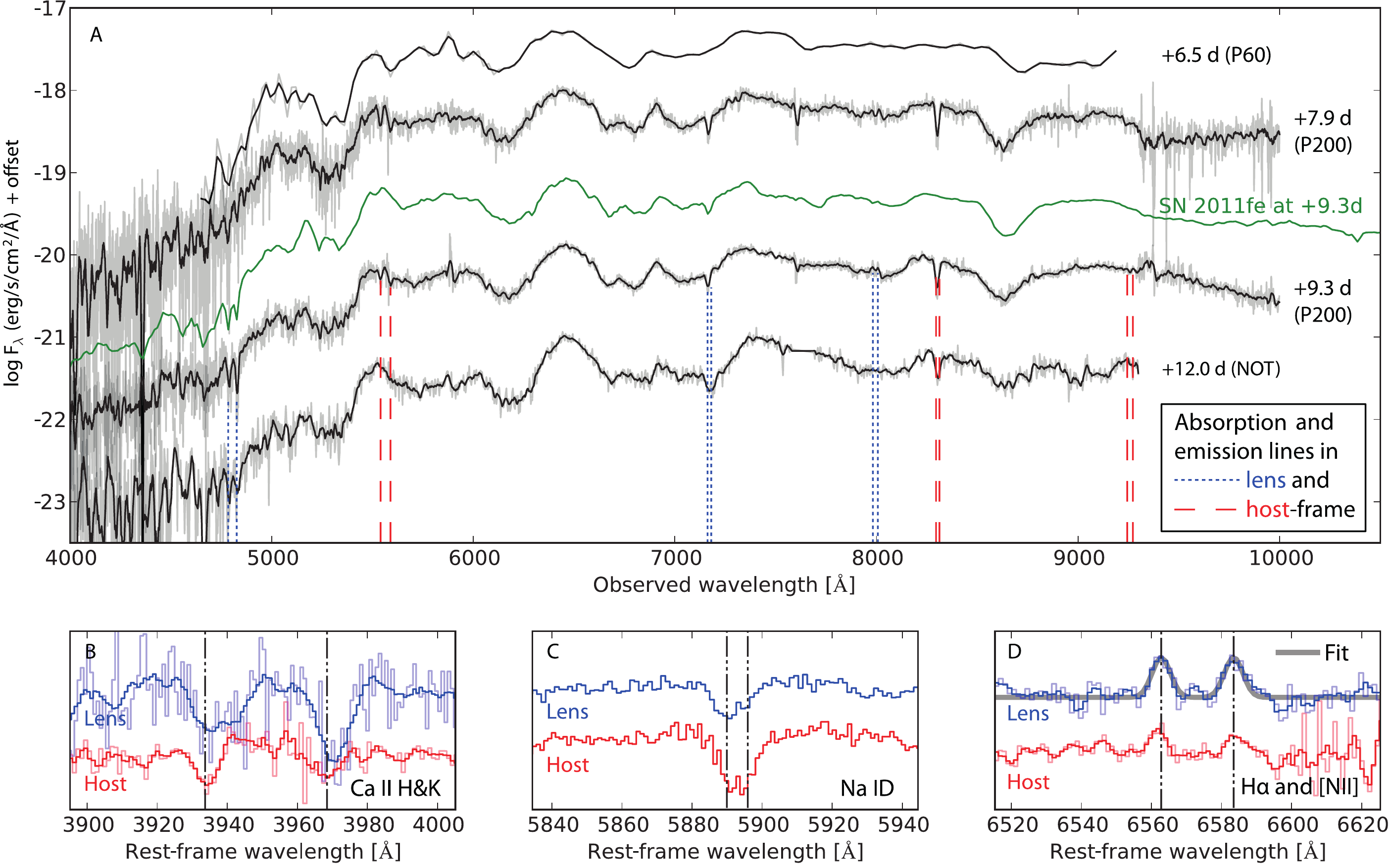}
	\caption{%
	{\bf Spectroscopic identification of  iPTF16geu as  a Type Ia supernova and measurements of the redshifts of the \sn host galaxy and the intervening lensing galaxy.} 
Measurements of the  \sn spectral energy distribution, $F_\lambda$,  obtained with the P60, P200 and NOT telescopes  
 are best fitted by a normal \snia spectral template.  Panel A shows a comparison with the near-by \snia, SN 2011fe, redshifted to $z=0.409$ (green line, \cite{2014MNRAS.439.1959M}) at
 a similar rest-frame phase, expressed in units of days with respect the time of optical lightcurve maximum.
  The spectra also reveal narrow absorption and emission lines, marked by the dashed vertical lines, from which the redshifts of the lens ($z=0.216$, blue lines) and \sn host galaxy ($z=0.409$, red lines) were determined. Zoomed in view in rest-frame wavelengths of the Ca~{\sc ii} H \& K and Na~{\sc i d} absorption features are shown in panels B, and C, respectively, together with the H$ {\alpha}$ and [N~{\sc ii}] emission lines (panel D). The H${\alpha}$ and [N~{\sc ii}] emission lines at $z=0.216$ were used to fit the velocity dispersion of matter in the lensing galaxy,  $\sigma_{v} = 163^{+41}_{-27}$ km s$^{-1}$.
\label{fig:spec}}
\end{figure}

Spectroscopic identification was carried out with the Spectral Energy Distribution (SED) Machine 
\cite{2014CoSka..43..209R} at the Palomar 60-inch telescope (P60) on 2016 October 2 and iPTF16geu was found to be spectroscopically consistent with a normal
\snia at $z\approx0.4$ (see Fig.~\ref{fig:spec}).  Further spectroscopic observations from the
Palomar~200-inch telescope (P200) and the 2.5-meter Nordic
Optical Telescope (NOT) were used to confirm the  \snia
identification and to establish the redshift of the host galaxy from
narrow sodium (Na~I~D) absorption lines, as $z=0.409$. The P200 and NOT spectra also
show absorption features from the foreground lensing galaxy at
$z=0.216$.  To estimate the velocity dispersion of the lensing galaxy, we fit two Gaussian functions with a common width to the H${\alpha}$ and [N~{\sc ii}] emission lines in the P200 spectrum in Fig~\ref{fig:spec}D. After taking the instrumental resolution into account, we measure $\sigma = 3.6^{+0.9}_{-0.6}$ \AA, corresponding to a velocity dispersion of $\sigma_{v} = 163^{+41}_{-27}$ km s$^{-1}$.

\begin{figure}[!htb]
	\centering
	\includegraphics[width=0.8\textwidth]{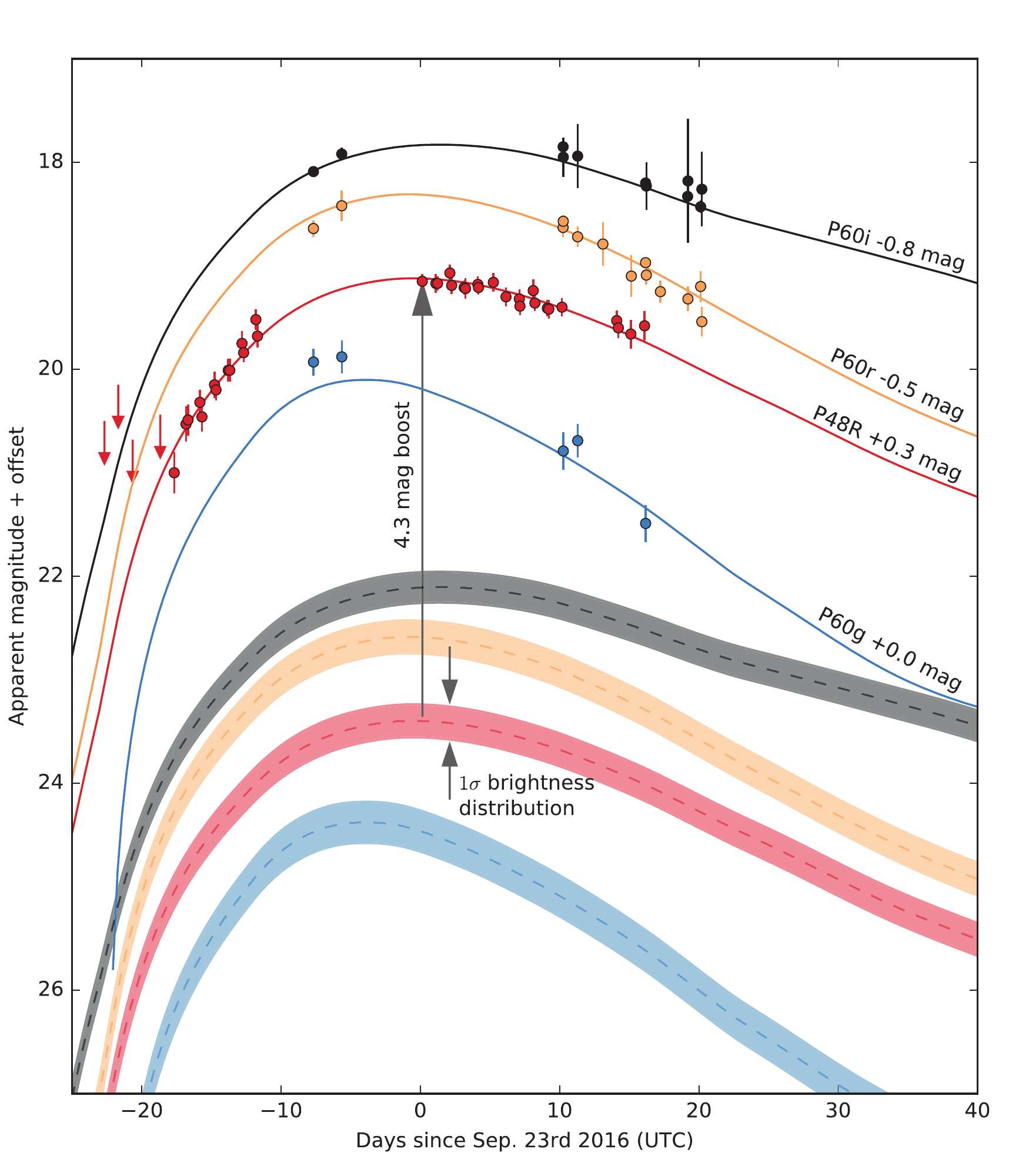}
	\caption{%
	{\bf Multi-color lightcurve of iPTF16geu showing that the supernova is 4.3 magnitudes (30 standard deviations) brighter than expected.}
	The magnitudes are measured with respect to time of maximum light (Modified Julian Date 57653.10) in $R$-band at P48 and $g$, $r$ and $i$-band with  RC on the
	SED Machine at P60. The filter transmission functions are shown in \cite{sup}.
		The solid lines show the best fitted SALT2 \cite{Guy:2007js} model to data.  The dashed lines indicate the 
		expected lightcurves at $z=0.409$ (without lensing) where the bands represent the standard deviation of the
		brightness distribution for \sneia.  In order to fit the observed lightcurves a brightness boost of 
		$4.3$~magnitudes is required.
		\label{fig:lc}}
\end{figure}

Photometric observations  of iPTF16geu collected at P48 and 
with the SED Machine Rainbow Camera (RC) at P60, between 2016 September 5 and October 13  (see Fig.~\ref{fig:lc}), were 
used to estimate the peak flux and lightcurve properties of the \sn with the SALT2 
lightcurve fitting tool \cite{Guy:2007js}.
The best fit lightcurve template, also shown in Fig.~\ref{fig:lc}, confirms that the observed 
lightcurve shapes are consistent with a \snia at $z=0.409$.  These fits also indicate 
some reddening of the supernova, suggesting that iPTF16geu suffers from moderate extinction 
by dust. This produces dimming at optical wavelengths of 20-40\%,  whith the largest losses 
in the $g$-band observations.  Thanks to the standard candle nature of \sneia, after correcting the peak magnitude for lightcurve properties \cite{1993ApJ...413L.105P,1998A&A...331..815T},
the flux of the \sn was found to be $\sim$30 standard deviations brighter than expected for the measured
redshift. 
This suggested that iPTF16geu was gravitationally lensed and we estimated the lensing amplification to be $\mu \sim 52$. Expressed in astronomical magnitudes, 
$\Delta  m = -4.3\pm {0.2}$~mag, where the uncertainty is dominated by the brightness dispersion of normal \sneia.
Since the magnification is derived from comparing the observed brightness of iPTF16geu to other 
\sneia \cite{Betoule:2014iz} within a narrow redshift range around $z=0.409$, the measurement of the lensing 
magnification is independent of any assumptions on cosmology, e.g., the value of the Hubble constant or 
other cosmological parameters. The lensing magnification is also independent of a lens model, which is the only 
way to determine the magnification for almost all other strong lensing systems.

\begin{figure}[!htb]
\centering
	\includegraphics[width=\textwidth]{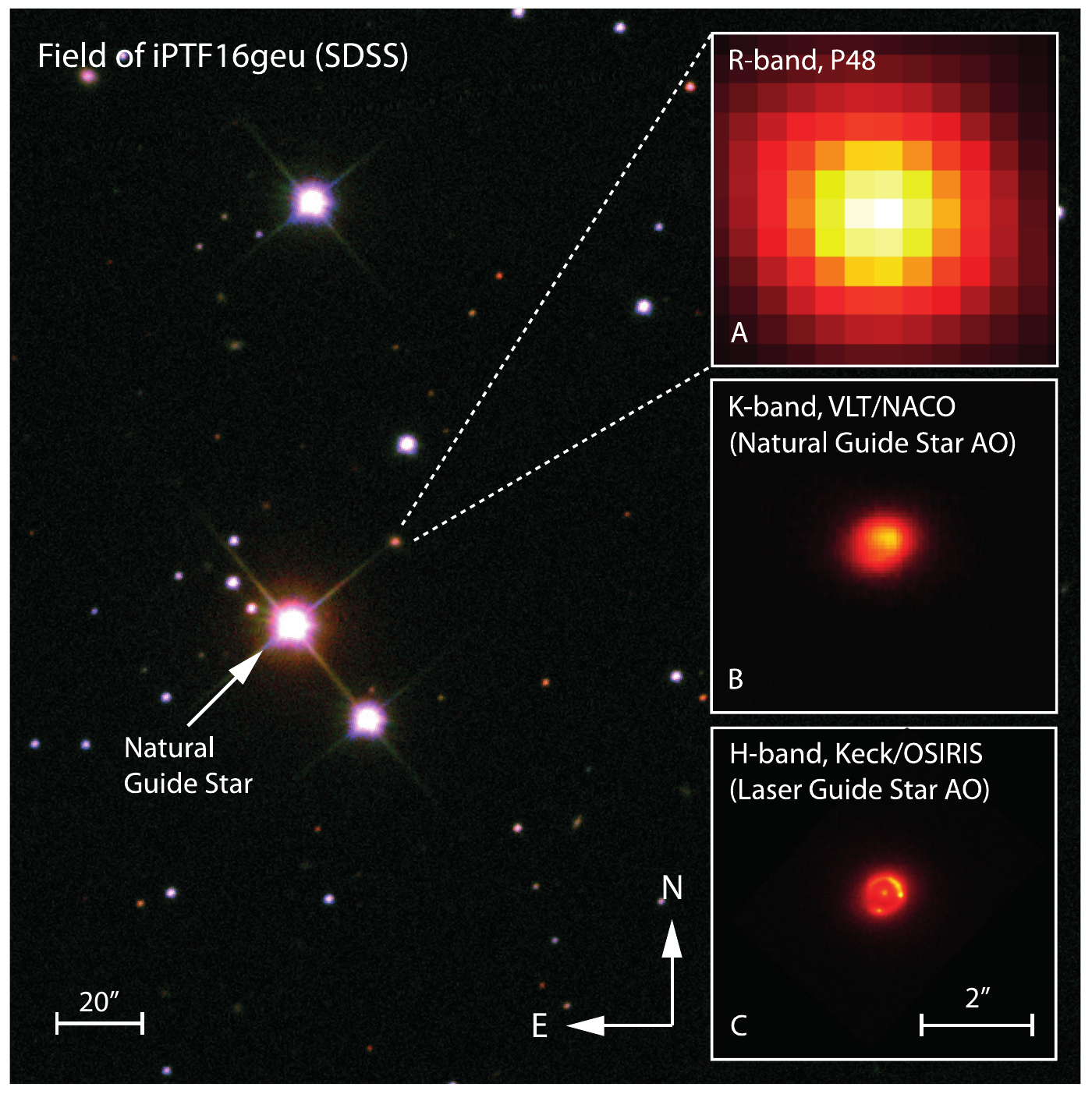}
	\caption{%
	{\bf Image of the field of iPTF16geu showing the spatial resolution of the ground based instruments used in this work.} A pre-explosion multicolor image from SDSS
	indicating the bright natural guide star and the position of the \sn detection in R-band at P48 (zoomed in panel A) near the galaxy SDSS J210415.89-062024.7.  The improved spatial resolution
	using a Natural star Guide System AO (NGSAO, panel B,  and further using the Laser Star aided AO System (LGSAO, panel C)  is shown.
		\label{fig:zoom}}
\end{figure}

The optical observations from Palomar, with a typical angular resolution (atmospheric seeing) of \ang{;;2},  were 
insufficient to spatially resolve any multiple images that could result from the strong lensing nature of the system 
(Fig.~\ref{fig:zoom}A). We therefore obtained $K_{\mathrm{s}}$-band (2.2\,$\mu$m) observations from the European Southern Observatory (ESO) with the Nasmyth 
Adaptive Optics System  Near-Infrared Imager and Spectrograph (NACO) at the Very Large Telescope (VLT). 
An angular resolution of $\sim$\ang{;;0.3} (full-width half-max, FWHM) was obtained at the location of the target. Adaptive optics (AO) 
corrections of the seeing were performed using a natural bright star,  $\sim$\ang{;;30} south-east of the \sn location,  
indicated in  Fig.~\ref{fig:zoom} along with the SDSS pre-explosion image of the field \cite{2015ApJS..219...12A}. 

The near-IR  image from VLT  indicated the structure expected in a  strongly lensed system, with higher flux in the  northeastern and southwestern regions of the system, compared to the center (Fig.~\ref{fig:zoom}B).
Multiple images of the system were first resolved with observations from the Keck observatory at near-infrared wavelengths, using the Laser Guide Star aided Adaptive Optics (LGSAO) with the
OH-Suppressing Infra-Red Imaging Spectrograph (OSIRIS) instrument, yielding an image quality of \ang{;;0.07}
FWHM in  the $H$-band  centered at 1.6 $\mu$m (Fig.~\ref{fig:zoom}C). 

\begin{figure}[!htb]
\centering
	\includegraphics[width=\textwidth]{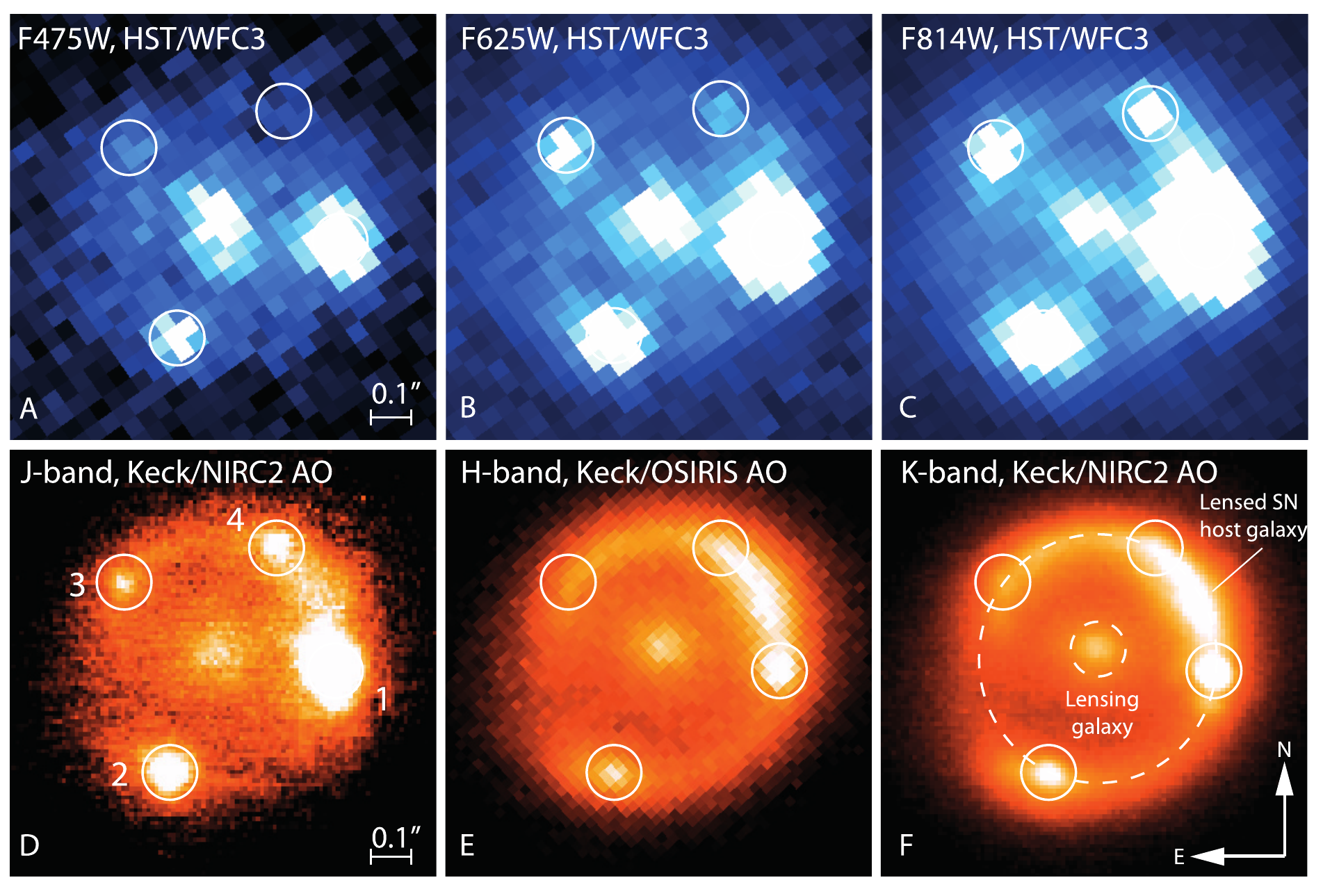}
	\caption{%
	{\bf High spatial resolution images from the Hubble Space Telescope and the Keck Observatory used to resolve the positions of the \sn images, the partial Einstein ring of the host galaxy and the
	intervening lensing galaxy.}
	HST/WFC3 observations of iPTF16geu obtained on 2016 October 25 in the $F475W$, $F625W$ and $F814W$ bands. are shown in the panels A,B and C respectively.  The images reveal four point sources,  except for $F475W$ where \sn images~3 and~4 are too faint. The NIR images obtained using Adaptive Optics aided Keck observations in the $J$, $H$ and $K_s$ bands are shown in panels
	D, E and F.  
	All four \sn images are clearly seen in $J$-band (panel D). For the $H$ and $K_s$ images, both the lensing galaxy at the center of the system and the lensed partial Einstein ring of the host galaxy are visible.  
		\label{fig:combo}}
\end{figure}

LGSAO observations of iPTF16geu using the Near-InfraRed Camera 2 (NIRC2) at the Keck telescope on 2016 October 22 and November 5,  in  $K_{\mathrm{s}}$-band and 
$J$-band (1.1$\mu$m) respectively, and optical images  obtained with the Hubble Space Telescope (HST) on 2016 October 25, are shown in Fig.~\ref{fig:combo}. The HST observations were carried out through the $F475W$, $F625W$ and
$F814W$ filters, where the names correspond to the approximate location of the central wavelength in nanometers.

The observations exhibit four images of  iPTF16geu, \ang{;;0.26}--\ang{;;0.31} from the
center of the lensing galaxy,  with nearly 90$^\circ$ azimuthal
separations. The extended host galaxy, warped by the lens to form a partial
Einstein ring, is brighter in the near-IR compared to the observations through optical filters. Thus,  the fainter individual \sn images are poorly resolved
for the observations with the longest wavelengths in Fig.~\ref{fig:combo}.  Furthermore, the \snia 
spectral energy distribution (redshifted to $z=0.4$) peaks within the $F625W$ and $F814W$ filters, see e.g. \cite{2014MNRAS.439.1959M}. 
Dimming by interstellar dust in the line of sight is roughly inversely proportional to wavelength in the optical and near-IR \cite{1989ApJ...345..245C}.
The biggest impact from extinction by dust is therefore expected for the shortest wavelength, in $F475W$ filter observations, where the two faintest \sn images cannot be detected above the
background light.
\newcommand{\saltc}{\mathcal{C}}
The low  spatial resolution lightcurves in Fig.~\ref{fig:lc} are  dominated by the two brightest \sn images, 
labelled 1 and 2 in Fig.~\ref{fig:combo}D. The $F625W$-$F814W$ magnitude difference (color) of the resolved images 
measured with HST indicate small differences in relative extinction between the \sn images, except for image 4, 
which appears to have about two magnitudes of additional dimming in $F814W$. 

Unaccounted dimming of light by scattering on dust grains in the line of sight would lead to an underestimation 
of the lensing amplification. Including corrections for differential extinction in the intervening lensing galaxy 
between the \sn images suggest a wider range for the lensing magnification of iPTF16geu, between $-4.1$ 
and $-4.8$ mag \cite{sup}.


The \sn multiple-image positions in Fig~\ref{fig:combo} were used to construct a lensing model,  with an isothermal ellipsoid galaxy
lens \cite{1993LIACo..31..571K,1994A&A...284..285K} with ellipticity $\epsilon_e=0.15\pm 0.07$ and mass
$M=(1.70 \pm 0.06)\cdot 10^{10}\,M_\odot$ inside an ellipse with major axis $1.13$ kpc and minor axis $0.97$ kpc.  Details of the lensing model are presented in the Supplementary Material \cite{sup}. The lens model can be independently verified through comparisons between the model-predicted and observed velocity dispersion of the lensing galaxy. From the model 
we derive an estimate,  $\sigma^{\rm mod}_v=156\pm 4$ km s$^{-1}$,  in good agreement with the measured value of the velocity dispersion (Fig.~\ref{fig:spec}D).

However, the adopted smooth isothermal ellipsoid lens model predicts brightness differences
between the multiple \sn images that are in disagreement with the observations. Including corrections for 
extinction in the resolved \sn images in the $F814W$ filter, we find large discrepancies between the model and 
measured magnitude differences for the multiple images of iPTF16geu:
   $\Delta m^{obs}_{1j}-\Delta m^{mod}_{1j}$  = ($-0.3,  -1.6, -1.5$) mag for $j=2,3$ and $4$, where the indices follow the numbering scheme adopted in Fig.~\ref{fig:combo}.
The observed discrepancy between the smooth model predictions for the  \sn images 1 and 2  compared to 3 and 4 (brighter by a factor 4 and 3, respectively),  cannot be accounted for by time-delays between the images, as they are 
predicted to be $<35$ hours \cite{sup}.  Graininess of the stellar distribution and dark matter sub-halos in the lens
galaxy, in addition to the smooth mass profile, can cause variations to magnification without altering image locations. 
These milli- and micro-lensing effects \cite{1989Natur.338..745K,1994ApJ...429...66W}, small enough not to cause additional resolved image separations, offer a plausible explanation for the deviation from the smooth lens model.

Available forecasts for wide-field surveys \cite{2010MNRAS.405.2579O} suggest that  about one strongly lensed \snia could be expected in our survey, 
irrespectively of redshift and magnification, with approximately a 30\% chance to be in a quad configuration.  For an average ellipticity of the lenses $e = 0.3$ \cite{2010MNRAS.405.2579O}, only about  1\% of the lensed \sne are expected to 
have $\mu \gsim 50$ \cite{Chae:2002uf}.
We have performed an independent rate estimate, with a somewhat simplified lensing simulation but including survey specific parameters, and confirm that the probability to detect and classify a
highly magnified \snia like iPTF16geu does not exceed the few percent level \cite{sup}.

  iPTF16geu appears to be a rather unlikely event, unless the actual rate of very magnified \sne is higher
  than anticipated, e.g., if the contribution from lensing by any kind of sub-structures in galaxies is underestimated, or if
  we are otherwise lacking an adequate description of gravitational lensing at the $\sim$ 1 kpc scale. The  physical scale probed by the resolved images of iPTF16geu is comparable to the smallest of the 299 multiply-imaged lensed systems in the Master 
Lens Database \cite{master}. Using the  standard candle nature of \sneia we can more easily detect strongly lensed systems with sub-arcsecond angular separations,
allowing exploration of the bending of light at  scales $\lsim$ 1 kpc, an otherwise challengingly small distance in studies of gravitational lensing \cite{2017ApJ...834L...5G}.  
As demonstrated with iPTF16geu, discovered while still brightening with a modest size telescope and sub-optimal atmospheric conditions, the locations of these rare systems can be identified
in advance of  extensive follow-up imaging at high spatial resolution. 
\bibliographystyle{Science}


\noindent AG and RA acknowledge support from the Swedish National Science Council (VR)
and the Swedish National Space Board. EM acknowledge support from VR. The iPTF Swedish collaboration is funded
through a grant from the K \& A Wallenberg foundation. We are grateful for
support from the National Science Foundation through the PIRE GROWTH project,
 Grant No 1545949.  PEN acknowledges support from the DOE under grant DE-AC02-05CH11231, Analytical Modeling for Extreme-Scale Computing Environments. This research used resources of the National Energy Research Scientific Computing Center, a DOE Office of Science User Facility supported by the Office of Science of the U.S. Department of Energy under Contract No. DE-AC02-05CH11231. MS acknowledges support from EU/FP7 ERC grant no. 615929.
 Some of the data presented herein were obtained at the W.M. Keck 
 Observatory, which is operated as a scientific partnership among the California Institute of 
 Technology, the University of California and the National Aeronautics and Space Administration. 
 The Observatory was made possible by the generous financial support of the W.M. Keck Foundation. 
 The data presented here were obtained in part with ALFOSC, which is 
 provided by the Instituto de Astrofisica de Andalucia (IAA) under a joint agreement with the 
 University of Copenhagen and NOTSA.  STSDAS and PyRAF are products of the Space 
 Telescope Science Institute, which is operated by AURA for NASA.  Part of the processing was 
 carried out off-line using the commercial software package MATLAB and Statistics Toolbox 
 Release 2013a, The MathWorks, Inc., Natick, Massachusetts, United States.  Some of the 
 data presented herein were obtained at the W.M. Keck Observatory, which is operated as a 
 scientific partnership among the California Institute of Technology, the University of California 
 and the National Aeronautics and Space Administration. The Observatory was made possible 
 by the generous financial support of the W.M. Keck Foundation. 
All data used in this paper are made public, including the photometric data (tables S1, S2, S5) and spectroscopic data at public repository WISeREP \cite{2012PASP..124..668Y} (http://wiserep.weizmann.ac.il). The
positions of the \sn images used in the lensing model are indicated in table S4.

\begin{flushleft}
{\bf SUPPLEMENTARY MATERIALS} \\
www.sciencemag.org \\
Materials and Methods \\
Figs. S1, S2, S3 \\
Tables S1, S2, S3, S4, S5 \\
References (34-56) \\
\end{flushleft}


\clearpage


\clearpage




\renewcommand{\theequation}{S\arabic{equation}}



\topmargin 0.0cm
\oddsidemargin 0.2cm
\textwidth 16cm 
\textheight 21cm
\footskip 1.0cm



\renewcommand\refname{References and Notes}



\centerline {\Large \bf Supplementary Materials} 

\newcommand{\geu}{iPTF16geu\xspace}
\section*{Materials and Methods}

\subsection*{Supernova survey and follow-up} 
Between late August and December 2016, the intermediate Palomar Transient Factory \cite{iptf} has
been performing a mixed-cadence experiment on the 48-inch Samuel Oschin Telescope (P48) at the 
Palomar Observatory searching for transient phenomena, e.g., supernovae. About 100 fields
($\simeq700$ square degrees in sky area) were observed every night, while another 200 fields were 
observed every three nights. During a night, each field was visited twice, with a one hour interval. For
fields with available SDSS $g$-and Mould $R$-band \cite{2012PASP..124..854O} reference images, 
these two visits were one in each filter. For fields with only $R$-band reference
images, these two visits were both in the $R$-band.

Images were transferred and processed by our real-time image subtraction pipeline at the National 
Energy Research Scientific Computing Center \cite{cnk16}. This pipeline, equipped with high-performance 
computing and machine-learning transient classifiers \cite{brp+13,rbw15}, delivered transient candidates 
for visual inspection by our duty astronomers within ten minutes of images being taken.

On 2016 Sep~11, our duty astronomer identified a transient candidate near the core of the galaxy 
SDSS\,J$210415.89$-$062024.7$ in one  $R$-band-only one-day cadence field. This candidate was 
saved as \geu (a.k.a. SN2016geu) and put in the queue for spectroscopic classification.  However,  
\geu was not observed spectroscopically until Oct~2, as described below, and both the redshift and the 
transient type were unknown until that point.

Meanwhile the field was observed with a daily cadence in the $R$-band, as listed in Table~\ref{tb:groundlog}.  
Difference imaging photometry in $R$-band was obtained using the iPTF Discovery Engine at the Infrared 
Processing and Analysis Center \cite[IDE]{2017PASP..129a4002M}, where references images are aligned and 
matched to have the same point spread function (PSF) to the images where the SN is active.  
PSF photometry is then carried out and calibrated by matching the photometry of the field stars to the 
stellar catalog from the Sloan Digital Sky Survey \cite[SDSS]{2015ApJS..219...12A}.

Complementary follow-up photometry, also listed in Table~\ref{tb:groundlog}, was obtained in the $g'r'i'$-bands 
with the Rainbow Camera (RC) on the Spectral Energy Distribution Machine \cite[SEDM]{2012SPIE.8446E..86B} mounted 
on the Palomar 60-inch telescope (P60). For the redshift of \geu, $z_{SN}=0.409$, these filters provide an 
excellent match to the  rest-frame $UBV$ filters (Fig.~\ref{fig:filters}) that have historically been 
used for studying \sneia. 
The RC data was pre-processed following standard photometry reduction techniques. The host subtraction 
was done by using the automatic reference-subtraction photometry pipeline \texttt{FPipe} 
\cite{2016A&A...593A..68F}.  This is using a similar approach to IDE, but images from the Sloan 
Digital Sky Survey are used as reference images of the host galaxy since such data is not available from the RC.
The photometric uncertainty is determined as the measured scatter from placing artificial 
PSFs in a circular pattern around the real transient and measuring their values on the subtracted images.
The P48 and P60 photometry is listed in Table~\ref{tb:palomarphot}.  
In the table, only the statistical uncertainty of each data point is quoted.  The systematic uncertainties 
are all subdominant in this analysis, since the error on the derived magnification will be 
dominated by the intrinsic \snia brightness uncertainty.

SEDM is a Caltech-developed instrument, designed for fast transient classification and follow-up. The 
SEDM focal plane combines a photometric instrument, the RC, and an Integral Field Unit (IFU) spectrograph. 
The focal plane of the RC is split into four quadrants, each one containing a SDSS generation 2 filter: 
$u'g'r'i'$. The field of view (FoV) is 6.5 arcmin$^2$ for each filter. The IFU, a lenslet array, is mounted 
in the center. The instrument covers the optical range 4000$-$9500$\,$\AA~with a constant wavelength 
resolving power of $R\simeq$100, equivalent to a velocity resolution of 3000$\,$km$\,$s$^{-1}$.  Its 
FoV is 30 arcsec$^2$ and each spaxel covers an approximate diameter of \ang{;;0.7}. A spaxel is a spatial 
pixel that also contains a spectrum of its detected photons, which is characteristic for an IFU.  

\subsection*{Spectroscopic follow-up}
The first classification spectrum of \geu was obtained with the IFU on the SEDM on
2016 Oct~2. 
The combined spectrum (Fig.~\ref{fig:spec}) consists of the average of two exposures, obtained with an 
offset of \ang{;;10} to  allow subtraction of the skylines.  The data were reduced using a custom IFU pipeline 
developed for the instrument .   Flux calibration and correction of telluric bands were done 
using the standard star {BD+28 4211}, which was taken at a similar airmass. An aperture of \ang{;;4} was used 
to extract the spaxels.  The details of all spectroscopic observations are listed in Table~\ref{tb:speclog}.


The classification and redshift of \geu from the low-resolution SEDM IFU spectrum was confirmed by 
observations obtained with the Double Spectrograph  \cite[DBSP]{1982PASP...94..586O} mounted on
the Palomar 200-inch telescope (P200) and the ALFOSC instrument on the Nordic Optical Telescope 
(NOT). 

For the DBSP observations the 600 lines/mm (with first order blaze wavelength 4000~\AA) grating was 
used with slitwidths \ang{;;1.5} and \ang{;;2} and the data were reduced with a custom pipeline written in the 
Interactive Data Language (IDL). Flux calibration and telluric correction were done with the flux standard star {BD+28 4211}.

For the ALFOSC spectrum grism \#4 was used with the slitwidth \ang{;;1.0}.  The data were reduced with a custom 
pipeline written in Matlab. Wavelength calibration was performed with a He-Ne arc lamp, and flux calibration with the 
standard star {BD+17 4708} observed the same night at similar airmass.

\subsection*{Adaptive Optics observations of \geu}
\subsubsection*{Natural Guide Star Adaptive Optics imaging from the VLT}
$K_{s}$-band observations were obtained on 2016~Oct~11 using the Nasmyth Adaptive Optics System with the Near-Infrared 
Imager and Spectrograph (NACO) at the Very Large Telescope (VLT).  
The bright star at R.A.=\ang{21;04;17.42} and Dec=\ang{-06;20;43.13}, \ang{;;29.5}$\,$ south-east of \geu with $V=$11.5$\,$mag 
marked in Fig.~\ref{fig:zoom}, was used as a natural guide star (NGS).  The spatial resolution decreases with the distance 
from the NGS, and \ang{;;30} is close to the maximum distance for this technique.  Due to problems with the visual 
wave-front sensor, the infrared wavefront sensor with the N20C80 dichroic was used. The observations were made 
with a standard jitter pattern in a \ang{;;5} box with single exposure integrations of 20\,s, saved in cube mode. In 
total, 348 frames were obtained. The data were reduced with the NACO \texttt{esorex} pipeline, using the standard 
jitter recipe \cite{2013A&A...559A..96F}. Since all exposures were individually saved, a small improvement in the final 
resolution could be obtained by selecting the best-resolved 20\,\% of the frames. The resulting image, Fig~\ref{fig:zoom}B, 
has an image quality of FWHM$\sim$0.3\arcsec\ at the location of the target, with a Strehl ratio of $\sim$5\%.  
The Strehl ratio is 
defined as the ratio of peak diffraction intensities of an aberrated and a perfect wavefront. The ratio indicates the level 
of image quality in the presence of wavefront aberrations.

\subsubsection*{Laser Guide Star Adaptive Optics imaging from Keck}\label{sec:keckao}
\geu was observed with the OSIRIS \cite{2006SPIE.6269E..1AL} imager behind the Keck~I Laser Guide Star 
Adaptive Optics (LGSAO) system on 2016~Oct~13 and separately with the Keck~II NIRC2 near-infrared imager in the 
$H$ and $K_s$-bands (at 1.6\,$\mu$m and 2.2\,$\mu$m respectively) on 2016 Oct~22 and~23, and then 
again in the $J$-band (at 1.2\,$\mu$m) on Nov 05.   The field of view of the OSIRIS imager is \ang{;;20}$\times$\ang{;;20} and is 
sampled with \ang{;;0.02} pixels by a 1024$\times$1024 Hawaii 1~HgCdTe array. The observation consisted of a sequence of 36 exposures of 
30\,s each in the OSIRIS ``$H_{\rm bb}$'' filter (1.638/0.330 $\mu$m), dithered in a 3$\times$3 box pattern 
with \ang{;;2.5} separation.  The NGS south-east of \geu mentioned above was used used for tip/tilt correction. Immediately 
following the \geu observation, the tip/tilt star was centered on the OSIRIS imager and a short sequence 
of dithered observations was taken as a PSF reference,  yielding point source FWHM$\simeq$\ang{;;0.07} 
(as measured from the $H = 15$ mag star at R.A.$=$\ang{21;04;18.042} and Dec$=$\ang{-06;20;43.19} 
near the bright tip/tilt star).

The OSIRIS data were reduced using standard infrared self-calibration
techniques. First, the 36 individual frames were scaled to a common
median and combined into a super-sky frame using a 3-$\sigma$ clipped
mean algorithm, masking out a \ang{;;2}$\times$\ang{;;2} box around the location
of the target source in each frame.  This super-sky was divided by its
own median in order to produce a super-flat-field image that was then used
to flat-field the individual science frames.  The median unmasked value
was then subtracted from each flat-fielded science frame in order to
produce the flat-fielded, foreground-subtracted science frames. These
frames were then shifted by the dither offsets
(adjusted slightly by hand to ensure proper alignment) and combined
using a 3-$\sigma$ clipped mean algorithm.  The final image is shown
in Figures~\ref{fig:zoom} and~\ref{fig:combo}.

The NIRC2 observations consist of 39 exposures of 80\,s each in the $K_s$-band,
9~exposures of 120\,s each in the $H$-band and 5~exposures of 300\,s each in the $J$-band. 
Unfortunately, for the $J$-band data, there was an instrument problem during the night and only two 
of the frames were used in the end. We used the same 
tip-tilt star as above to do the AO corrections. The observations were carried out using 
the NIRC2 narrow camera with a field of view of \ang{;;100}$\times$\ang{;;100} and a pixel scale of \ang{;;0.01}/pixel. 
We dithered each of the frames by \ang{;;2} using a custom nine-point dithering pattern for better sky 
subtraction. To correct for flat-fielding and dark currents we acquired a set of ten dark frames 
and twenty dome flat frames. The dome flats were separated into two sets with the first half with the dome flat 
lamp off and second half with the lamp on. The dome flat off frames were used to estimate the thermal radiation 
which is non-negligible for the $K_s$-band.  The dome flat on frames where then subtracted with the combined
flat-off map, and the individual dark and flat frames where combined to generate master dark and dome flat frames, 
respectively. 
This produced a combined dark-subtracted and flat-fielded images in the $J$, $H$ and $K_s$ bands.  
For each science exposure the sky background of each pixel were estimated by using the preceding and following 
images after bad pixels and the object it self had been masked.  The reduction was carried out using custom python
scripts.  The images were then finally corrected for geometric distortion using IDL routines 
available on the NIRC2 webpage. 

The NIRC2 $J$-band, Fig.~\ref{fig:combo}D, provides the highest resolution image in 
our dataset of the system where the four SN images are visible. We used this image to determine the SN 
positions by fitting a model to the system.  The lensing galaxy was modeled using a 
S\'ersic profile \cite{1963BAAA....6...41S} while Gaussians were used for the the SN images.  The fitted SN
positions are shown in Table~\ref{tb:snpos}.

\subsection*{Observations with the Wide-Field Camera~3 aboard the Hubble Space Telescope}
The Hubble Space Telescope (HST) Wide Field Camera~3 (WFC3) images presented in Fig.~\ref{fig:combo} A--C were 
obtained under program DD~14862 (Principal Investigator: Goobar) using the ultraviolet and visual (UVIS) channel on 
2016~Oct~24 in the $F475W$, $F625W$ and $F814W$ filters (where the filter names correspond to the central 
wavelengths in nm). UVIS consists of two thinned, backside illuminated, ultra violet optimized 2048$\times$4096 pixel CCDs.  
However, only part of the UVIS2 chip was read for the \geu observations using the \texttt{UVIS2-C512C-SUB} aperture.   
UVIS has a pixel scale of \ang{;;0.04}/pixel and the diffraction limited PSF for these filters results in an image 
quality of FWHM$\simeq$\ang{;;0.07}. We used a standard 3-point dithering pattern with post-flash to maximize 
the charge transfer efficiency (CTE) during read-out.  The total exposure time for the three filters were 
$378\,$s, $291\,$s and $312\,$s for $F475W$, $F625W$ and $F814W$, respectively. 

The images were automatically processed at the Space Telescope Science Institute first through version~3.3 of the 
\texttt{calwf3} pipeline where they are dark subtracted, flat-fielded and corrected for charge-transfer inefficiency.  
Further, the individual flatfielded images were combined and corrected for geometric distortion using the
\texttt{AstroDrizzle} software.

Using the same model as for the SN image positions, we fitted the relative fluxes of SN images 2--4 with respect to 
image 1.  For this procedure we kept the image positions fixed to the values in Table~\ref{tb:snpos}, while allowing for a 
shift and a rotation of the the whole model.  For the $F625W$ band, which covers a similar wavelength range as the 
$R$-band, we found that 90\,\% of the total flux is contained in images 1--2 and 70\,\% of the total flux is contained in 
image~1 alone.  The relative brightnesses for the SN images with respect to the brightest image~1, the relative 
reddening, and the extinction these correspond to assuming the extinction law from \cite{1989ApJ...345..245C} are 
shown in Table~\ref{tb:hstres}.

\subsection*{Fitting to a \snia lightcurve template}
The lightcurves from the P48 and P60 photometry are presented in Figs.~\ref{fig:lc}, and~\ref{fig:lcw11fe}.  
Since \sneia are a homogenous class of objects, templates or lightcurve models can typically be fitted to 
the data. The free parameters are the time, $t_0$, and brightness, $m_B^\star$, of maximum in the rest-frame 
$B$-band, the lightcurve width and the color of the SN.  

The lightcurve width can be quantified by introducing a stretch factor, $s$, that scales the time 
variable of the template with respect to $t_0$.  It has been shown that same behavior is also captured by 
the second principal component of the SALT2 lightcurve model \cite{Guy:2007js}.  In short, the SALT2 model
is constructed from a principal component analysis of a large data set of normal \sneia.  The eigenvalue for the
first component is directly related to the peak brightness, $m_B^\star$, while the eigenvalue for the 
second component, $x_1$, is normalized so that a SN with $x_1=0$ correspond to an average normal SN Ia. 
Further, objects with $x_1=\pm1$ have lightcurve widths that correspond to $\pm1\sigma$ of normal \sneia 
lightcurve width distribution.

The color is typically defined in terms of the rest-frame $B-V$ magnitude.  This is then used to scale a 
color, or extinction, law that describes how the flux ratios vary with wavelength.  For a standard extinction law, the color excess,
$E(B-V)$, is used to scale the law, where $E(B-V)$ is the $B-V$ magnitude deviation from
what would have been measured in the absence of extinction. SALT2 uses an empirically 
derived color law where the scaling parameter, $\saltc$, is the deviation from the average
$B-V$ rest-frame color.

We fitted both a lightcurve template and the SALT2 model to the P48 and P60 photometry.
In Fig.~\ref{fig:lc} we show the best fitted SALT2 model.  SALT2 was also used for the 
\snia cosmology sample presented in \cite{Betoule:2014iz}.  Using the same model allows us to 
estimate the magnification independently of the value of the Hubble constant, $H_0$, or any 
other cosmological parameter.  The SALT2 model has four free parameters, for which we obtain,  
$t_0=57654.1\pm0.2$, $m_B^{\star}=19.12\pm0.03$,  and $\saltc=0.23\pm0.03$ and 
$x_1=0.08\pm0.19$, respectively.   
The corrected peak magnitude, $m_B^\mathrm{corr}$, is further obtained as
\[
	m_B^\mathrm{corr} = m_B^{\star} - \beta\cdot\saltc + \alpha\cdot x_1
\]
where $\beta=3.101\pm0.075$ and $\alpha=0.141\pm0.006$ have been derived simultaneously with the cosmological
parameters by \cite{Betoule:2014iz} from the full sample of 740 \sneia.  Since their data set is well-sampled around 
$z=0.4$, we can compare the lightcurve of \geu with the expected average for the same redshift. This is shown in 
Fig.~\ref{fig:lc} together with the intrinsic brightness dispersions of \sneia.  By then comparing the derived peak 
magnitude of \geu with the expected for an unlensed \snia at $z=0.409$, we find that the SN has been boosted 
by $-4.3\pm0.2$~magnitudes where the intrinsic dispersion is accounted for in the quoted error bar.

We also tried to fit the spectral series of SN~2011fe, which is a normal and well observed \snia, to the data. 
In Fig.~\ref{fig:lcw11fe} we show both the SALT2 lightcurve model and the best SN~2011fe fit using the spectral 
series for the latter compiled by \cite{Amanullah:2015bj}.  The data are perfectly consistent with both 
models but SN~2011fe provides a better fit in the rest-frame  $U$-band.  However, since the SALT2 fit allows us 
to directly compare the brightness of \geu to other \sneia at the same redshift in a cosmology independent 
manner we decided to use this for the main analysis.   iPTF16geu has the same lightcurve shape as SN2011fe, 
but a reddening of $E(B-V)=0.31\pm0.05$~mag is required, assuming the extinction law from
\cite{1989ApJ...345..245C}.
 
Both the SALT2 model and SN~2011fe fits show that the \geu is consistent with a normal and 
average \sneia, but the derived values of $\saltc$ and $E(B-V)$ suggest that that the \sn is 
reddened by interstellar extinction.  This reddening is taken into account for the estimated 
lensing magnification and uncertainty above.  However, given that most of the measured light, 
$\sim70\,\%$, is contained in SN image~1, the measured $\saltc$ is dominated by the color of 
this image.  Since we have also measured differential reddening of images 2--4 with respect to 
SN image~1, the magnification of the system could in fact be higher.  If the measured differential 
extinction from Table~\ref{tb:hstres} is taken into account we find that the total magnification 
of the system lie in the range $-4.1$~mag to $-4.8$~mag.



\subsection*{Modelling of the gravitational lens}
Our default lens model is an isothermal ellipsoid galaxy \cite{1993LIACo..31..571K,1994A&A...284..285K} 
for which the convergence is given by
\be
\kappa =\frac{b}{2\omega}, 
\ee
where $b$ is a convergence normalization parameter and
\be
\omega^2=(1-\epsilon)(x-x_0)^2+(1+\epsilon)(y-y_0)^2.
\ee
Here, $x_0$ and $y_0$ correspond to the position of the lens centre and $(1+\epsilon)/(1-\epsilon)$ is the major-to-minor axis ratio.
The magnification is given by \cite{1994A&A...284..285K} 
\be
\mu (x_1,x_2)=\left(1-\frac{b}{\omega}\right )^{-1},
\ee
i.e., the isodensity contour $\kappa=1/2$ correspond to the critical line in the lens plane. In the forthcoming, positions in the lens plane are given in angular units. In terms of $\epsilon$, the ellipticity, $\epsilon_e$, of the galaxy is given by
\be
\epsilon_e=1-\sqrt{\frac{1-\epsilon}{1+\epsilon}}.
\ee
The projected mass $M$ inside an isodensity contour of constant $\omega$ is given by \cite{1994A&A...284..285K}
\be
M=\frac{c^2}{4G}\frac{D_sD_l}{D_{ls}}\frac{b\omega}{\sqrt{1-\epsilon^2}}.
\ee
Assuming cosmological parameter values of $H_0=67.8$ km s$^{-1}$Mpc$^{-1}$ and $\om=0.308$ and zero spatial curvature \cite{Ade:2015xua},
$$D_l=745.8\, {\rm Mpc}, D_s=1157.0\, {\rm Mpc}, {\rm and} \,D_{ls}=513.2\, {\rm Mpc},$$
where $D_l$ is the angular diameter distance to the lens, $D_s$ to the source and $D_{ls}$ between the lens and the source.
The mass inside the critical line where $\omega=b$ is 
\be
M=\left(\frac{b}{1''}\right)^2\frac{2.0\cdot 10^{11}\,M_\odot}{\sqrt{1-\epsilon^2}}.
\ee
Using the {\tt lensmodel} software \cite{Keeton:2001sr,Keeton:2001ss}, we fit the model to the
observed SN positions by varying $b$, $\epsilon$, the position angle $\phi$ (from from North to East) of 
the major axis, and the offset, $(\Delta \alpha,\Delta \delta)$, between the center of the lens model 
and the S\'ersic profile used to fit the observed light of the lensing galaxy as described above.  
The results are
\begin{eqnarray*}
b&=&\ang{;;0.287}\pm\ang{;;0.005}\,,\\
\Delta \alpha &=&\ang{;;-0.013} \pm \ang{;;0.007}\,,\\
\Delta \delta &=&\ang{;;-0.004} \pm \ang{;;0.008}\,,\\
\epsilon&=&0.16 \pm  0.09\,,\\
\phi&=&65.8 \pm  0.9\,,
\end{eqnarray*}
or 
\begin{eqnarray}
M&=&(1.69 \pm 0.06)\cdot 10^{10}\,M_\odot,\\
\epsilon_e&=&0.15 \pm  0.07.
\end{eqnarray}
For the best fit parameter values we get $\chi^2_{\rm min}=0.022$. Constraining the centre 
of the lens to coincide with the maximum surface brightness of the lens galaxy, i.e. $(\Delta \alpha,\Delta \delta) = (0,0)$ within the observational uncertainties, i.e., $\pm$ \ang{;;0.008}, the minimum 
$\chi^2$ increases to $\chi_{\rm min}^2=1.8$.  Following \cite{Chae:2002uf}, we calculate the velocity 
dispersion by averaging over all possible inclination angles and intrinsic axis ratios that can give 
rise to the observed projected surface density. The resulting relation between $b$ and the velocity 
dispersion, $\sigma_v$, is given by 
\be
b=4\pi\left(\frac{\sigma_v^2}{c^2}\right)^2\frac{D_{ls}}{D_s}\sqrt{1-\epsilon}\lambda \left(\sqrt{\frac{1-\epsilon}{1+\epsilon}}\right),
\ee
where the functional form of $\lambda$ depends on the probability for the three dimensional mass distribution of the lens to be oblate or prolate. Assuming total ignorance of this probability, we can read off the value of $\lambda$ for $\epsilon=0.16$ from figure~1 in \cite{Chae:2002uf} to be $\lambda=1\pm0.05$. Numerically, we have
\begin{eqnarray*}
\sigma_v&=280\left(\frac{b}{\lambda}\sqrt{1-\epsilon}\right)^\frac{1}{2} \, \, {\rm km\,s}^{-1} &=156\pm 4\,{\rm km\,s}^{-1},
\end{eqnarray*}
where $\lambda\sqrt{1-\epsilon}=0.92\pm 0.04$. 

We calculate the expected flux ratios, $r$, and their associated uncertainties by taking a weighted average of the flux ratios over a grid of parameter values, here denoted by index $i$, 
\begin{eqnarray}
\bar r &=& \sum_i{{\cal L}_ir_i},\\
\sigma^2_{\bar r}&=&\sum_i{{\cal L}_i(r_i-\bar r)^2},
\end{eqnarray}
where the weight is given by the likelihood of the parameter values given the observed image positions, 
\be
{\cal L}_i\equiv\exp\left(-\frac{1}{2}\chi_i^2\right)/\sum_i{\exp\left(-\frac{1}{2}\chi_i^2\right)}.
\ee
Suppressing the bar as an indicator for average values, the magnitude differences $\Delta m_{ij}=-2.5\ln{f_i/f_j}$ between images $i$ and $j$ are:
\begin{center}
\begin{tabular}{lll}
$\Delta m_{12} = -0.39\pm 0.11$, &  $\Delta m_{13} =-0.33\pm 0.06$, & $\Delta m_{14}=0.73\pm 0.06$,\\
$\Delta m_{23} =  \, \, \, \, \, 0.04\pm 0.11$, &  $\Delta m_{24} =\, \, \, \,\, 1.11\pm 0.06$, & $\Delta m_{34} =1.06\pm 0.06$.
\end{tabular}
\end{center}
In conclusion, image 4 is expected to be the brightest, while it is  in fact observed to be the faintest. 

The predicted time delay, calculated as weighted averages analogously to the flux ratios, between the \sn images is small,  ranging from $1.9\pm 1.2$ hours to $15.7\pm 6.3$ hours. Thus, the maximal time-delay between any two 
images is 35 hours at the 99.9\% confidence level with the adopted model.

While the model estimates of the magnitude differences among the \sn
images have small uncertainties, the total magnification is poorly
constrained. Since there is a tail of very high magnification output
compatible with the observed image positions, the resulting weighted
mean total magnification is very sensitive to details of the grid of
parameter values and possible cuts of low probability parameter
values. Only when disregarding all parameter values not being within
2$\sigma$ of the best fit parameters can the total magnification be
constrained at all, $\mu_{\rm tot}=43\pm 29$, consistent with the
observed magnification $\mu\sim 52$.

Motivated by the discrepancy between the image magnitude differences
observed and the ones predicted by the smooth lensing model, we
consider the possibility of sub-structures as possible added
contributors to the lensing of one or more \sn images.

Such sub-structures can include galactic subhalos 
\cite{Springel:2008cc}, and compact objects in the form of stars or
possible compact dark matter objects. The magnification of SNe from
stars is studied in \cite{Dobler:2006wv}, where it is found that
$\sim 70\,\%$ of multiple lensed SNe will experience additional
magnification from the lens galaxy star population of $>0.5$ mag. The
case of lensed quasars is studied in, e.g \cite{2014ApJ...793...96S}.

A population of compact objects will create a network of caustics that
can give rise to large magifications if the source size is
sufficiently small. If there is a relative motion between the source,
observer and the lensing compact object, or if the source size is
changing, the magnification can vary. The time scale of the variations
will typically correspond to the Einstein radius crossing time (see
below), but can also be shorter in the case of caustic crossing events
\cite{Wambsganss:2006nj}.

A full treatment of the gravitational lensing effects of
sub-structures is beyond the scope of this paper, but in order to show
the typical scales involved, we briefly study the lensing effect of a
single compact object.

In the isolated point mass lens approximation, the relevant source plane length scale is given by \cite{1992grle.book.....S}
\be
\eta_0 = \sqrt{2R_S\frac{D_sD_{ls}}{D_l}}=\sqrt{\frac{M}{M_\odot}}\times 3.8\cdot 10^{16}\,{\rm cm},
\ee
where $R_S$ is the Schwarzschild radius of the lens. To act efficiently as a lens, $\eta_0$ should be larger than the physical size, $R$, of 
the Type Ia SNe photosphere, typically
\be
\frac{R}{10^{15}\,{\rm cm}}\sim 2.0+0.62\frac{(t-t_{\rm max})}{1\,{\rm week}}\,,
\ee
which is derived from the the measured expansion velocities of the photosphere of SN2011fe \cite{2013A&A...554A..27P,2014MNRAS.439.1959M}.

A single lens point mass should have $M\gsim 0.12 M_\odot$ to be effective also one month after maximum. 
For a lens with velocity 156 km s$^{-1}$, this corresponds to time scales greater than 9 years. We can thus neglect any
time dependence induced to the lightcurve from the motion of isolated point masses with $M \gsim 0.12 M_\odot$.

\subsection*{Predicted rate of highly magnified \sneia in iPTF}
To date, almost 2000 \sneia up to $z \sim 0.2$ have been discovered at P48 and
  classified over a period of eight years (945 observing nights), with a detection limit of $R \sim 21$
  mag. Since they are brighter, lensed \sne can be observed to higher redshifts.
  Using the compilation of  measured \snia volumetric rates in \cite[{\rm see their Fig.~2}]{2011ARNPS..61..251G}, we estimate that within the larger
  volume up to $z=0.4$, the number of \sneia explosions is 12.6
  deg$^{-2}$ year$^{-1}$, which when combined with the total monitoring time 
  and the average solid angle of the survey, yields a total of $6
  \times 10^{4}$ \sneia in the field of view. This corresponds to a relative fraction
  $1.7 \times 10^{-5}$ for events like iPTF16geu, if this single event
  is representative for the true rate. Estimates of the inefficiencies in transient detection and spectroscopic typing over the lifetime of the survey
  have been included in the calculation.
  
  To compare this outcome with the estimate of the expected probability we use the ray-tracing SNOC Monte-Carlo package \cite{2002A&A...392..757G}. SNOC was
  used in \cite{2002A&A...393...25G} to assess the expected rate of strongly lensed \sne in planned satellite missions. 
       We run a total
     of $10^7$ simulated lines-of-sight up to $z=0.45$, with a volumetric redshift dependent \snia rate following the star formation rate with redshift, one
     of the options in SNOC.      
     The matter spatial distribution in galaxy halos along the line of sight are simulated with two different smooth spherical functions, the 
     Singular Isothermal Sphere (SIS) profile,
     $$ \rho(r) = \frac{\sigma^2_v}{2\pi G r^2}, $$
     where $\sigma^2_v$ is the velocity dispersion; and the NFW profile \cite{1997ApJ...490..493N}, 
    $$\rho(r) = \frac{\rho_0}{ \frac{r}{R_s} \left(1 + \frac{r}{R_s}\right)^2},$$
     where $\rho_0$ and $R_s$ parameters of the model and vary between halos. Since we only
     consider spherical profiles, our simulations  will not produce four, but only two images.  Furthermore, these simulations do not include the contribution from microlensing by a stellar population in the line of sight. 

 However, in addition to the smooth spherical functions, we tested adding contributions
      from point-like lenses (POI) to the mass density. In all cases, the contributions to the mass density is self-consistently normalised to give the 
      cosmological average mass density of the universe found by the Planck collaboration for a  $\Lambda$CDM universe, $\Omega_M=0.308$ \cite{Ade:2015xua}.  
     The resulting distribution of the expected  gravitational lensing amplification for the iPTF survey, $\mu$, can be seen in Fig.~\ref{fig:snoc}.
      For the simulations with smooth distributions, e.g., NFW,  no event like iPTF16geu came out of the simulations, corresponding to a probability less than $10^{-3}$ . Adding substructures
      in the from of compact objects increases the chances of intersecting a lens, yet the high magnification found for iPTF16geu is
      rare. For a 90\% contribution of compact objects to the cosmic mass density, $\Omega_M $, we find a 5\% chance of discovering a \snia with comparable
      lensing magnification within the limited redshift range probed. In conclusion, the lensing characteristics of iPTF16geu are rather unlikely, even considering the possibility of 
      having a high density of discrete compact sources.


\renewcommand{\thetable}{S\arabic{table}}
\setcounter{table}{0}
\begin{center}
{\footnotesize
\begin{longtable}{llccc}
\caption[Imaging of iPTF16geu.]{%
  {\bf Imaging of iPTF16geu.}  All observations where imaging data of iPTF16geu were obtained.
} \label{tb:groundlog} \\
\hline\hline
\multicolumn{1}{c}{UTC Civil date} & \multicolumn{1}{c}{MJD} & Telescope & Filter & Exp. time\\
& (days) & & & (s)\\
\hline
\endfirsthead

\hline
\multicolumn{1}{c}{UTC Civil date} & \multicolumn{1}{c}{MJD} & Telescope & Filter & Exp. time\\
& (days) & & & (s)\\
\hline
\endhead

\hline 
\multicolumn{5}{r}{{\footnotesize{\it Continued on next page}}} \\ \hline
\endfoot

\hline \hline
\endlastfoot
2016 Sep 05.3 & 57636.33 & P48 & $R$ & 60 \\
2016 Sep 06.2 & 57637.18 & P48 & $R$ & 120 \\
2016 Sep 07.2 & 57638.18 & P48 & $R$ & 120 \\
2016 Sep 08.2 & 57639.23 & P48 & $R$ & 120 \\
2016 Sep 09.2 & 57640.21 & P48 & $R$ & 120 \\
2016 Sep 10.2 & 57641.20 & P48 & $R$ & 120 \\
2016 Sep 11.2 & 57642.20 & P48 & $R$ & 120 \\
2016 Sep 15.3 & 57646.33 & P60/RC & $r'$ & 90 \\
2016 Sep 15.3 & 57646.33 & P60/RC & $i'$ & 90 \\
2016 Sep 15.3 & 57646.33 & P60/RC & $g'$ & 90 \\
2016 Sep 17.4 & 57648.36 & P60/RC & $r'$ & 90 \\
2016 Sep 17.4 & 57648.36 & P60/RC & $i'$ & 90 \\
2016 Sep 17.4 & 57648.36 & P60/RC & $g'$ & 90 \\
2016 Sep 23.1 & 57654.14 & P48 & $R$ & 60 \\
2016 Sep 24.1 & 57655.11 & P48 & $R$ & 120 \\
2016 Sep 25.1 & 57656.12 & P48 & $R$ & 120 \\
2016 Sep 26.1 & 57657.15 & P48 & $R$ & 120 \\
2016 Sep 27.1 & 57658.13 & P48 & $R$ & 120 \\
2016 Sep 28.2 & 57659.25 & P48 & $R$ & 60 \\
2016 Sep 29.1 & 57660.15 & P48 & $R$ & 60 \\
2016 Sep 30.1 & 57661.11 & P48 & $R$ & 120 \\
2016 Oct 01.1 & 57662.10 & P48 & $R$ & 120 \\
2016 Oct 02.1 & 57663.12 & P48 & $R$ & 120 \\
2016 Oct 03.2 & 57664.17 & P48 & $R$ & 60 \\
2016 Oct 03.2 & 57664.25 & P60/RC & $r'$ & 90 \\
2016 Oct 03.3 & 57664.25 & P60/RC & $i'$ & 90 \\
2016 Oct 03.3 & 57664.26 & P60/RC & $r'$ & 90 \\
2016 Oct 03.3 & 57664.26 & P60/RC & $i'$ & 90 \\
2016 Oct 03.3 & 57664.26 & P60/RC & $g'$ & 90 \\
2016 Oct 04.3 & 57665.29 & P60/RC & $r'$ & 90 \\
2016 Oct 04.3 & 57665.30 & P60/RC & $i'$ & 90 \\
2016 Oct 04.3 & 57665.30 & P60/RC & $g'$ & 90 \\
2016 Oct 06.1 & 57667.11 & P60/RC & $r'$ & 90 \\
2016 Oct 07.1 & 57668.10 & P48 & $R$ & 120 \\
2016 Oct 08.1 & 57669.11 & P48 & $R$ & 60 \\
2016 Oct 08.1 & 57669.15 & P60/RC & $r'$ & 90 \\
2016 Oct 09.1 & 57670.10 & P48 & $R$ & 60 \\
2016 Oct 09.2 & 57670.17 & P60/RC & $r'$ & 90 \\
2016 Oct 09.2 & 57670.17 & P60/RC & $i'$ & 90 \\
2016 Oct 09.2 & 57670.17 & P60/RC & $g'$ & 90 \\
2016 Oct 09.2 & 57670.22 & P60/RC & $r'$ & 90 \\
2016 Oct 09.2 & 57670.23 & P60/RC & $i'$ & 90 \\
2016 Oct 10.2 & 57671.22 & P60/RC & $i'$ & 90 \\
2016 Oct 10.2 & 57671.24 & P60/RC & $r'$ & 90 \\
2016 Oct 10.2 & 57671.24 & P60/RC & $i'$ & 90 \\
2016 Oct 11.1 & 57672.05 & VLT/NACO & $K_s$ & 6960\\
2016 Oct 12.2 & 57673.19 & P60/RC & $r'$ & 90 \\
2016 Oct 12.2 & 57673.19 & P60/RC & $i'$ & 90 \\
2016 Oct 12.2 & 57673.21 & P60/RC & $r'$ & 90 \\
2016 Oct 12.2 & 57673.21 & P60/RC & $i'$ & 90 \\
2016 Oct 13.1 & 57674.12 & P60/RC & $r'$ & 90 \\
2016 Oct 13.1 & 57674.12 & P60/RC & $i'$ & 90 \\
2016 Oct 13.2 & 57674.21 & Keck/OSIRIS & $H_{\rm bb}$ & 1080\\
2016 Oct 13.2 & 57674.21 & P60/RC & $r'$ & 90 \\
2016 Oct 13.2 & 57674.21 & P60/RC & $i'$ & 90 \\
2016 Oct 22.2 & 57683.22 & Keck/NIRC2 & $K_s$ & 3120\\
2016 Oct 23.2 & 57684.22 & Keck/NIRC2 & $H$ & 1080\\
2016 Oct 24.9 & 57685.88 & HST/WFC3 & $F625W$ & 291\\
2016 Oct 24.9 & 57685.89 & HST/WFC3 & $F814W$ & 312\\
2016 Oct 24.9 & 57685.92 & HST/WFC3 & $F475W$ & 378\\
2016 Nov 05.2 & 57697.21 & Keck/NIRC2 & $J$ & 600\\
\end{longtable}}

\end{center}

\begin{center}
{\footnotesize
\begin{longtable}{lrccrrc}
\caption[Measured ground-based optical photometry of iPTF16geu from Palomar Observatory.]{%
{\bf Photometry of individual exposures of iPTF16geu from the P48 and P60/RC observations
listed in Table~\ref{tb:groundlog}.  The photometry is given in flux, $f$, which can be converted to magnitudes, $m$,
in the AB system as $m=-2.5\log_{10}(f) + ZP$, with $ZP=25$. \label{tb:palomarphot}}}  \\
\hline\hline
\multicolumn{1}{c}{UTC Civil date} & \multicolumn{1}{c}{MJD (days)} & Telescope & Filter & \multicolumn{1}{c}{Flux (arbitrary)} & \multicolumn{1}{c}{Flux $\sigma$} & ZP (AB)\\
\hline
\endfirsthead

\hline
\multicolumn{1}{c}{UTC Civil date} & \multicolumn{1}{c}{MJD (days)} & Telescope & Filter & \multicolumn{1}{c}{Flux (arbitrary)} & \multicolumn{1}{c}{Flux $\sigma$} & ZP (AB)\\
\hline
\endhead

\hline \multicolumn{7}{r}{{\footnotesize{\it Continued on next page}}} \\ \hline
\endfoot

\hline\hline
\endlastfoot
2016 Aug 31.3 & 57631.33 & P48 & $R$ & 19 & 10 & 25.0\\
2016 Sep 01.3 & 57632.32 & P48 & $R$ &  8 & 13 & 25.0\\
2016 Sep 02.4 & 57633.35 & P48 & $R$ & 17 &  8 & 25.0\\
2016 Sep 04.3 & 57635.33 & P48 & $R$ & 38 & 10 & 25.0\\
2016 Sep 05.3 & 57636.33 & P48 & $R$ & 52 & 10 & 25.0 \\
2016 Sep 06.2 & 57637.18 & P48 & $R$ & 81 & 13 & 25.0 \\
2016 Sep 06.3 & 57637.33 & P48 & $R$ & 84 & 12 & 25.0 \\
2016 Sep 07.2 & 57638.18 & P48 & $R$ & 98 & 11 & 25.0 \\
2016 Sep 07.3 & 57638.33 & P48 & $R$ & 86 & 11 & 25.0 \\
2016 Sep 08.2 & 57639.23 & P48 & $R$ & 115 & 14 & 25.0 \\
2016 Sep 08.3 & 57639.34 & P48 & $R$ & 110 & 10 & 25.0 \\
2016 Sep 09.2 & 57640.21 & P48 & $R$ & 131 & 13 & 25.0 \\
2016 Sep 09.3 & 57640.32 & P48 & $R$ & 131 & 13 & 25.0 \\
2016 Sep 10.2 & 57641.20 & P48 & $R$ & 166 & 18 & 25.0 \\
2016 Sep 10.3 & 57641.32 & P48 & $R$ & 153 & 13 & 25.0 \\
2016 Sep 11.2 & 57642.20 & P48 & $R$ & 205 & 19 & 25.0 \\
2016 Sep 11.3 & 57642.31 & P48 & $R$ & 177 & 18 & 25.0 \\
2016 Sep 15.3 & 57646.33 & P60/RC & $r'$ & 221 & 16 & 25.0 \\
2016 Sep 15.3 & 57646.33 & P60/RC & $i'$ & 278 & 10 & 25.0 \\
2016 Sep 15.3 & 57646.33 & P60/RC & $g'$ & 107 & 13 & 25.0 \\
2016 Sep 17.4 & 57648.36 & P60/RC & $r'$ & 270 & 37 & 25.0 \\
2016 Sep 17.4 & 57648.36 & P60/RC & $i'$ & 325 & 18 & 25.0 \\
2016 Sep 17.4 & 57648.36 & P60/RC & $g'$ & 112 & 16 & 25.0 \\
2016 Sep 23.1 & 57654.14 & P48 & $R$ & 288 & 19 & 25.0 \\
2016 Sep 24.1 & 57655.11 & P48 & $R$ & 283 & 23 & 25.0 \\
2016 Sep 24.2 & 57655.24 & P48 & $R$ & 283 & 18 & 25.0 \\
2016 Sep 25.1 & 57656.12 & P48 & $R$ & 310 & 23 & 25.0 \\
2016 Sep 25.2 & 57656.25 & P48 & $R$ & 278 & 20 & 25.0 \\
2016 Sep 26.1 & 57657.15 & P48 & $R$ & 273 & 18 & 25.0 \\
2016 Sep 26.2 & 57657.25 & P48 & $R$ & 270 & 25 & 25.0 \\
2016 Sep 27.1 & 57658.13 & P48 & $R$ & 281 & 21 & 25.0 \\
2016 Sep 27.2 & 57658.18 & P48 & $R$ & 273 & 18 & 25.0 \\
2016 Sep 28.2 & 57659.25 & P48 & $R$ & 286 & 24 & 25.0 \\
2016 Sep 29.1 & 57660.15 & P48 & $R$ & 251 & 21 & 25.0 \\
2016 Sep 30.1 & 57661.11 & P48 & $R$ & 247 & 20 & 25.0 \\
2016 Sep 30.2 & 57661.15 & P48 & $R$ & 231 & 19 & 25.0 \\
2016 Oct 01.1 & 57662.10 & P48 & $R$ & 265 & 27 & 25.0 \\
2016 Oct 01.2 & 57662.23 & P48 & $R$ & 238 & 18 & 25.0 \\
2016 Oct 02.1 & 57663.12 & P48 & $R$ & 227 & 17 & 25.0 \\
2016 Oct 02.2 & 57663.22 & P48 & $R$ & 225 & 19 & 25.0 \\
2016 Oct 03.2 & 57664.17 & P48 & $R$ & 229 & 19 & 25.0 \\
2016 Oct 03.2 & 57664.25 & P60/RC & $r'$ & 223 & 18 & 25.0 \\
2016 Oct 03.3 & 57664.25 & P60/RC & $i'$ & 347 & 29 & 25.0 \\
2016 Oct 03.3 & 57664.26 & P60/RC & $r'$ & 236 & 13 & 25.0 \\
2016 Oct 03.3 & 57664.26 & P60/RC & $i'$ & 316 & 55 & 25.0 \\
2016 Oct 03.3 & 57664.26 & P60/RC & $g'$ & 48 & 8 & 25.0 \\
2016 Oct 04.3 & 57665.29 & P60/RC & $r'$ & 205 & 19 & 25.0 \\
2016 Oct 04.3 & 57665.30 & P60/RC & $i'$ & 319 & 91 & 25.0 \\
2016 Oct 04.3 & 57665.30 & P60/RC & $g'$ & 53 & 8 & 25.0 \\
2016 Oct 06.1 & 57667.11 & P60/RC & $r'$ & 192 & 37 & 25.0 \\
2016 Oct 07.1 & 57668.10 & P48 & $R$ & 203 & 19 & 25.0 \\
2016 Oct 07.2 & 57668.22 & P48 & $R$ & 191 & 18 & 25.0 \\
2016 Oct 08.1 & 57669.11 & P48 & $R$ & 180 & 23 & 25.0 \\
2016 Oct 08.1 & 57669.15 & P60/RC & $r'$ & 145 & 27 & 25.0 \\
2016 Oct 09.1 & 57670.10 & P48 & $R$ & 194 & 25 & 25.0 \\
2016 Oct 09.2 & 57670.17 & P60/RC & $r'$ & 163 & 6 & 25.0 \\
2016 Oct 09.2 & 57670.17 & P60/RC & $i'$ & 251 & 12 & 25.0 \\
2016 Oct 09.2 & 57670.17 & P60/RC & $g'$ & 25 & 4 & 25.0 \\
2016 Oct 09.2 & 57670.22 & P60/RC & $r'$ & 146 & 12 & 25.0 \\
2016 Oct 09.2 & 57670.23 & P60/RC & $i'$ & 244 & 52 & 25.0 \\
2016 Oct 10.2 & 57671.22 & P60/RC & $i'$ & 227 & 368 & 25.0 \\
2016 Oct 10.2 & 57671.24 & P60/RC & $r'$ & 126 & 13 & 25.0 \\
2016 Oct 10.2 & 57671.24 & P60/RC & $i'$ & 242 & 187 & 25.0 \\
2016 Oct 12.2 & 57673.19 & P60/RC & $r'$ & 120 & 135 & 25.0 \\
2016 Oct 12.2 & 57673.19 & P60/RC & $i'$ & 223 & 29 & 25.0 \\
2016 Oct 12.2 & 57673.21 & P60/RC & $r'$ & 118 & 13 & 25.0 \\
2016 Oct 12.2 & 57673.21 & P60/RC & $i'$ & 256 & 141 & 25.0 \\
2016 Oct 13.1 & 57674.12 & P60/RC & $r'$ & 132 & 18 & 25.0 \\
2016 Oct 13.1 & 57674.12 & P60/RC & $i'$ & 203 & 11 & 25.0 \\
2016 Oct 13.2 & 57674.21 & P60/RC & $r'$ & 96 & 12 & 25.0 \\
2016 Oct 13.2 & 57674.21 & P60/RC & $i'$ & 238 & 79 & 25.0 \\
\end{longtable}}

\end{center}

\begin{center}
\begin{table}[!htb]
\begin{center}
{\footnotesize
\caption{%
  {\bf Spectroscopic observations of iPTF16geu used for identification and redshift measurements.}
 	\label{tb:speclog}}
\begin{tabular}{lccccccccc}
\hline\hline
UT Date &  MJD   & Phase & Telescope  & $R$ & $\Delta \lambda$ 
 & $\lambda$ range & Exp. time  &  Airmass \\ 
& (days) & (days) & Instrument &  ($\lambda / \Delta \lambda$) & (\AA) & (\AA) & (s) \\ \hline
Oct 2.23        & 57663.23      & 6.5   & P60/SEDM                        & 100   & 58    & 4000-9500             & 2700                & 1.43 \\
Oct 4.22        & 57665.22      & 7.9   & P200/DBSP             & 740   & 8               & 3300-10000     & 1800  & 1.37 \\
Oct 6.13        & 57667.13      & 9.3   & P200/DBSP             & 560   & 10    & 3300-10000    &  3600  & 1.51 \\
Oct 9.90        & 57670.90      & 12.0           & NOT/ALFOSC            & 360   & 16    & 3500-9600             & 2700             & 1.22 \\
\hline
\end{tabular}}
\end{center}
\end{table}
\end{center}

\begin{table}[!htb]
\begin{center}
\caption{%
{\bf Fitted relative SN image positions.}  
The positions are given in polar coordinates with respect to the center, 
of the S\'ersic profile, and the angle, $-\pi<\phi\leq\pi$, is defined from West to North.  
The SN image numbers are shown in Fig.~\ref{fig:combo}.  The fitted statistical uncertainty 
on $r$ and  are $\phi$ \ang{;;0.001} and 0.004~rad, respectively.  The uncertainty, in the 
position of the whole system, i.e. the center of the S\'ersic profile is \ang{;;0.008}
in both $x$ and $y$.
\label{tb:snpos}}
\begin{tabular}{c c c}
\hline\hline
SN Image & $r$ & $\phi$\\
& (arcsec) & (rad)\\
\hline
1 &  0.2679 & $+1.7861$\\
2 &  0.3133 & $-2.7252$\\
3 &  0.2874 & $-0.9761$\\
4 &  0.2803 & $+0.4554$\\
\hline
\end{tabular}
\end{center}
\end{table}

\begin{table}[!htb]
\begin{center}
\caption{%
  {\bf Relative photometry with respect to the first SN image}.  Here, $E(F625W-F814W)$, is the relative color color excess with respect to SN image 1.  All quoted uncertainties are statistical errors.  In the last two columns, the relative extinction are given using the measured $E(F625W-F814W)$ together with the extinction law from \cite{1989ApJ...345..245C}.
\label{tb:hstres}}
\begin{tabular}{cccccc}
\hline\hline
SN image & $\Delta F625W$ & $\Delta F814W$ & $E(F625W-F814W)$ & $A_{F625W}$ & $A_{F814W}$\\
 & (mag) & (mag) & (mag) & (mag) & (mag) \\
\hline
2 & $1.26 (0.01)$ & $1.12 (0.01)$ & $0.14 (0.01)$ & $0.5$ & $0.4$\\
3 & $2.46 (0.02)$ & $2.33 (0.02)$ & $0.12 (0.03)$ & $0.5$ & $0.4$\\
4 & $3.62 (0.05)$ & $2.88 (0.03)$ & $0.75 (0.06)$ & $2.8$ & $2.1$\\
\hline
\end{tabular}
\end{center}
\end{table}

\clearpage

\renewcommand{\thefigure}{S\arabic{figure}}
\setcounter{figure}{0}

\begin{figure}[!htb]
\centering
\includegraphics[width=\textwidth]{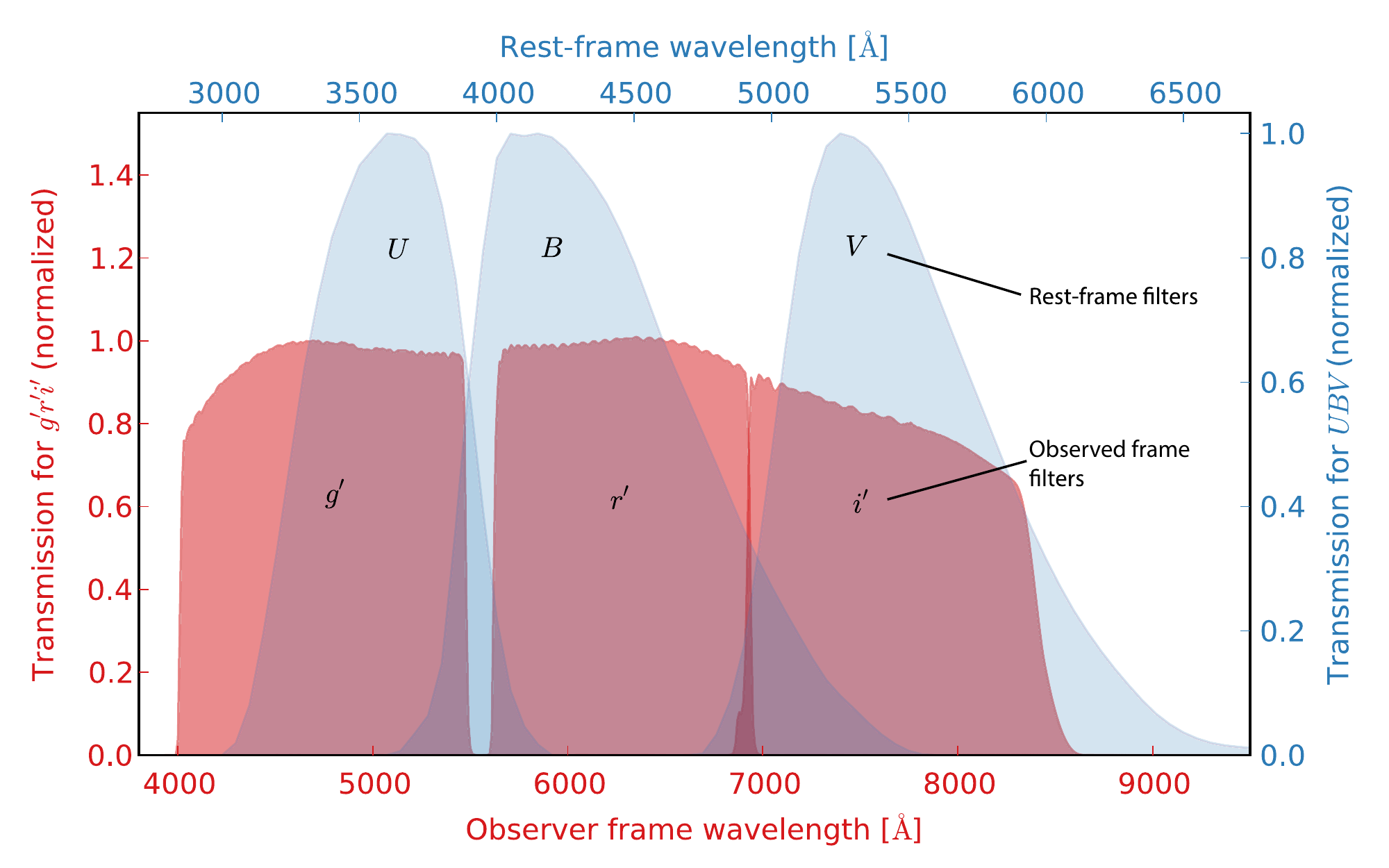}
\caption{%
{\bf Comparison between the observer frame P60 RC filters and the rest-frame standard filters $UBV$, 
which historically have been used for studying \sneia.}.  
The relative transmissions of the observer frame P60 $g'r'i'$ filters are shown in red for the wavelengths 
indicated by the bottom horizontal axis, while the corresponding rest-frame wavelengths and $UBV$ filter transmissions are plotted in blue.  It is clear from the figure that the observer frame RC $g'r'i'$ filters provide a close match to 
the rest-frame $UBV$ for the redshift $z_{SN}=0.409$.\label{fig:filters}}
\end{figure}

\begin{figure}[!htb]
\centering
\includegraphics[width=\textwidth]{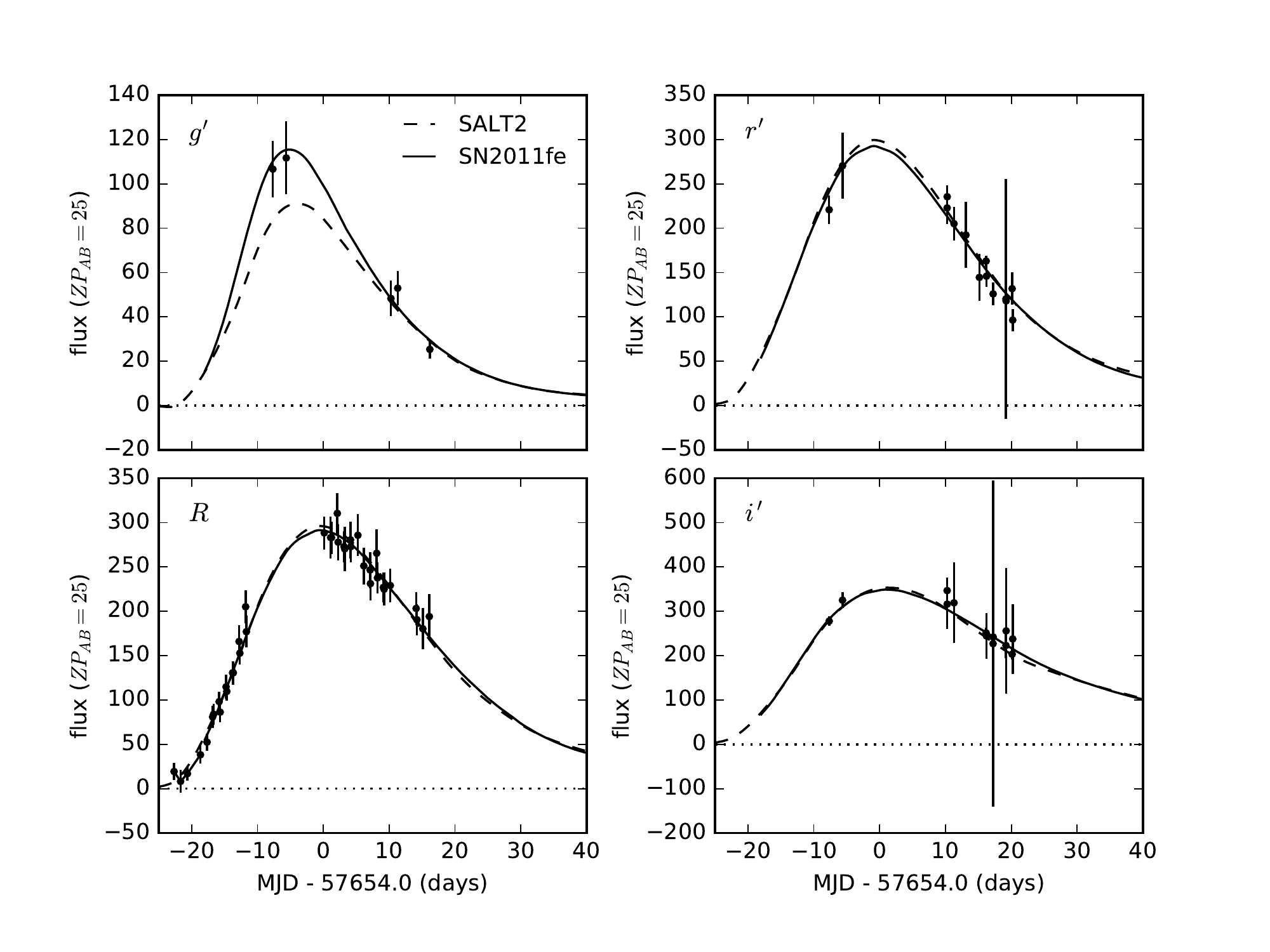}
\caption{%
{\bf Lightcurves of \geu in the P48 $R$-band and the P60 RC $g'r'i'$ bands.} The error bars are correlated.  
The solid lines show the fit of the nearby and normal SN~2011fe, while the dashed show the best fit
using the SALT2 model.
The lightcurve fitting is carried out in linear flux space which appropriately allows the inclusion of low 
signal-to-noise and negative flux measurements.  These points have been omitted in the plotting of 
Fig.~\ref{fig:lc}, which is shows magnitudes, while all data are shown here.  The individual fluxes are 
given in Table~\ref{tb:palomarphot}.
\label{fig:lcw11fe}}
\end{figure}

\begin{figure}[!htb]
\centering
\includegraphics[width=\textwidth]{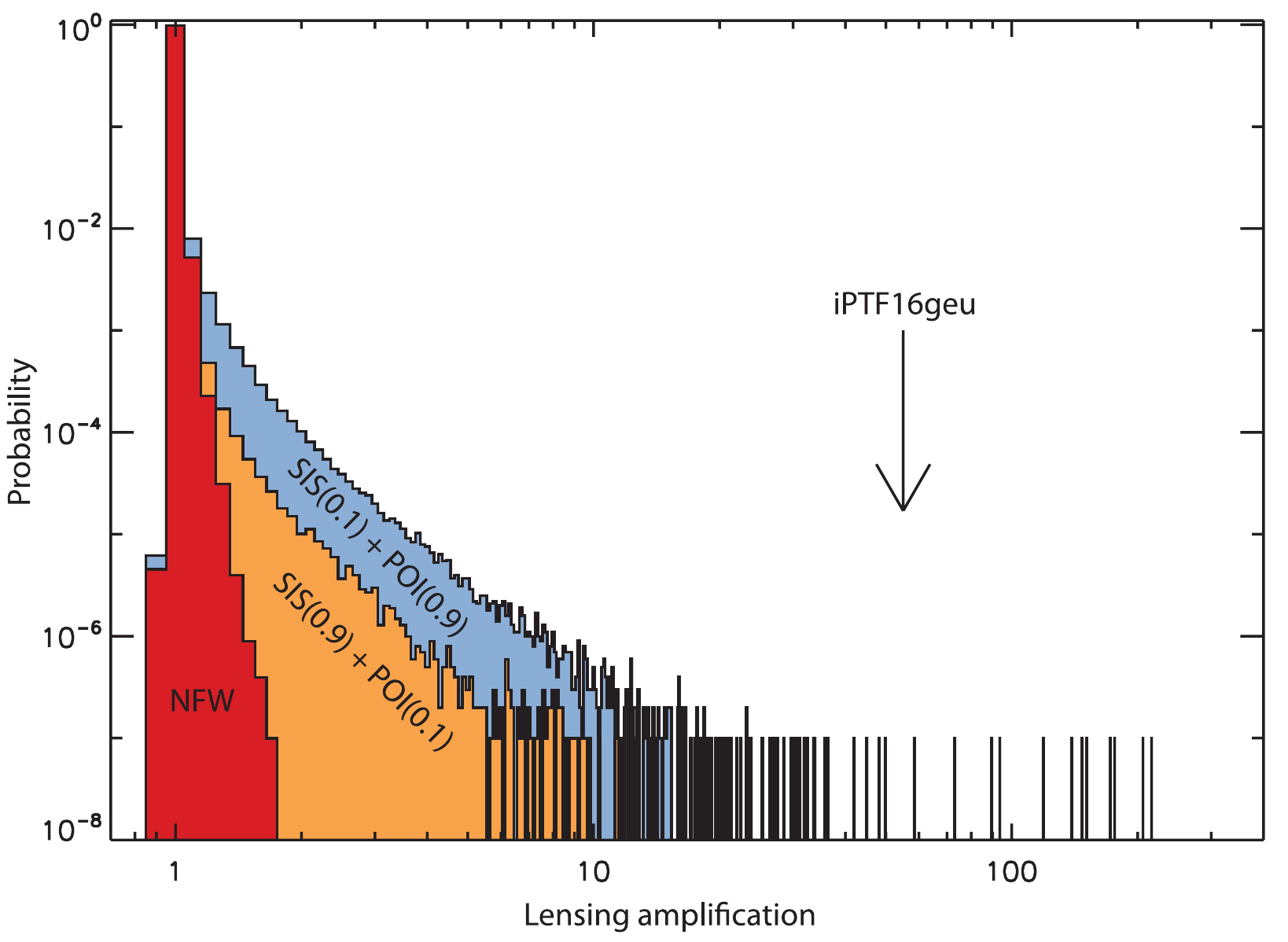}
\caption{%
{\bf Computation of the lensing probability for \sneia in the Palomar Transient Factory survey.} SNOC simulations \cite{2002A&A...392..757G} of the 
expected  likelihood for the gravitational lensing amplification of \sneia up to $z=0.45$ as a function if the density distribution of matter
in galaxy halos. The three cases displayed correspond to a smooth NFW profile, and a SIS profile with a 10\% and 90\% fraction of 
substructures in the form of point-like lenses (POI). Only simulations with substructures provide non-vanishing probability of finding an event with as 
high amplification as iPTF16geu, indicted by the arrow.}
\label{fig:snoc}
\end{figure}


\end{document}